\documentclass[10pt,showpacs,prd,amsmath,two column,amssymb]{revtex4}
\usepackage{amssymb}
\usepackage{txfonts}
\usepackage{amsfonts}
\usepackage{epsfig,bm,dcolumn}
\usepackage{graphicx}
\usepackage{color}
\usepackage{slashed}
\usepackage{overpic}

\begin{document}
\title{Effects of chiral imbalance and magnetic field on pion superfluidity and color superconductivity}
\author{Gaoqing Cao and Pengfei Zhuang}
\affiliation{ Department of Physics, Tsinghua University and Collaborative Innovation Center of Quantum Matter, Beijing 100084, China}
\date{\today}

\begin{abstract}
The effects of chiral imbalance and external magnetic field on pion superfluidity and color superconductivity are investigated in extended Nambu--Jona-Lasinio models. We take Schwinger approach to treat the interaction between charged pion condensate and magnetic field at finite isospin density and include simultaneously the chiral imbalance and magnetic field at finite baryon density. For the superfluidity, the chiral imbalance and magnetic field lead to catalysis and inverse catalysis effects, respectively. For the superconductivity, the chiral imbalance enhances the critical baryon density, and the magnetic field results in a de Haas--van Alphan oscillation on the phase transition line.
\end{abstract}

\pacs{21.65.Qr, 11.30.Rd, 05.30.Fk}

\maketitle

\section{Introduction}
\label{section1}

Quantum systems in an external magnetic field exhibit various unexpected and interesting features. In condensed matter physics, quantum hall effect and fractional quantum hall effect were discovered in two-dimensional electron systems in 1970s and 1980s~\cite{ando,klitzing,tsui}. More recently, quantum spin hall effect was proposed to exist in special two-dimensional electron systems with strong spin-orbit couplings~\cite{kane,bernevig}, and quantum anomalous hall effect was found in a ferromagnetic material chromium-doped (Bi,Sb)2Te3 in 2013~\cite{chang}. For a dense non-relativistic system, the de Haas--van Alphan oscillation was observed in some thermodynamic systems~\cite{haas,landau}. In high energy physics, this oscillation is found in nuclear and quark matters~\cite{ebert,noronha,fukushima,fayazbakhsh}. It is beyond our expectation that the magnetic field reduces the critical temperature of chiral symmetry restoration in Quantum Chromodynamics (QCD), known as the inverse magnetic catalysis~\cite{bali,bali2,bali3,bali4,bali5}.

In high energy heavy ion collisions, there might exist charge asymmetry on the opposite sides of the reaction plane, induced by the chiral magnetic effect~\cite{adler,bell,kharzeev,kharzeev2,skokov}. Besides a very strong magnetic field (up to $eB\sim 10m_\pi^2$ at the LHC energy) created in the early stage of heavy ion collisions, the randomly generated chiral imbalance plays also an important role in the hot medium. In neutron stars where the magnetic field and baryon chemical potential are both very large, chiral current can be induced along the magnetic field by triangle anomalies~\cite{son,metlitski,xuguang}. The chiral current will eventually cause chiral separation and generate chiral imbalance which can be quite large in the two hemispheres along the magnetic field. While the chiral imbalance or instead chiral chemical potential $\mu_5$ together with the magnetic field $B$ created by the spectators have been widely investigated in the study of chiral symmetry restoration~\cite{fukushima2,chao,cao} at finite temperature, it is still an open question how the chiral imbalance and magnetic field affect pion superfluidity at finite isospin density and color superconductivity at finite baryon density.

The pion superfluidity in an external magnetic field was studied in Nambu--Jona-Lasinio (NJL) model~\cite{kang}, linear sigma model~\cite{loewe} and Lattice QCD~\cite{endrodi}. When charged pion condensate exists, $u$ and $d$ quark fields are no longer the eigenstates of the quark propagator in flavor space, and a direct extension from the case without magnetic field $B$ to the case with finite $B$ by substituting the transverse momenta with the Landau levels of $u$ and $d$ quarks is invalid. The model calculation at hadron level shows a magnetic catalysis which is inconsistent with the lattice simulation at quark level. The color superconductivity in an external magnetic field is not so complicated as pion superfluidity. While the color condensates are also charged, the effective charges of $u$ and $d$ quarks with red (green) and green (red) colors are opposite to each other~\cite{fayazbakhsh}, when one takes the interaction between quarks and massless gluons into account. Thus, the color condensates can be effectively treated as neutral in an external magnetic field.

We investigate in this paper both the chiral imbalance and magnetic field effects on pion superfluidity and color superconductivity in extended NJL models. Considering the difficulty to treat the charged pion condensate in a magnetic field, we study separately the chiral imbalance effect and magnetic field effect, and for the latter we employ the Schwinger approach to include the interaction between the
charged pion condensate and the magnetic field. For the color superconductivity, we take into account the chiral imbalance and magnetic field simultaneously.

\section{pion superfluidity with chiral imbalance}
\label{section2}

The Lagrangian density of the extended NJL model with quark isospin chemical potential $\mu_I$ and chiral chemical potential $\mu_5$ is defined as
\begin{equation}
{\cal L}=\bar\psi\left(i\slashed\partial-m_0+{\mu_I\over2}\gamma_0\tau_3+\mu_5\gamma_0\gamma_5\right)\psi
+{G}\left[\left(\bar\psi\psi\right)^2+\left(\bar\psi i\gamma_5{\boldsymbol \tau}\psi\right)^2\right],
\end{equation}
where $\psi=\left(u,d\right)^T$ is the two-flavor quark field, $\tau_i$ are pauli matrices in flavor space, $m_0$ is the current quark mass, and $G$ is the coupling constant with dimension GeV$^{-2}$. In order to study the bound states of the system, we introduce four auxiliary fields $\sigma=-2G\bar\psi\psi$ and ${\boldsymbol \pi}=-2G\bar\psi i\gamma_5{\boldsymbol \tau}\psi$, and the Lagrangian density becomes
\begin{eqnarray}
{\cal L} &=& \bar\psi\bigg[i{\slashed \partial}-m_0-\sigma-i\gamma_5\left(\tau_3\pi_0+\tau_\pm\pi_\pm\right)+{\mu_I\over2}\gamma_0\tau_3\nonumber\\
&&+\mu_5\gamma_0\gamma_5\bigg]\psi-{\sigma^2+\pi_0^2+\pi_\mp\pi_\pm\over 4G},
\end{eqnarray}
where the auxiliary fields are related to the physical fields $\sigma$, $\pi_0=\pi_3$ and $\pi_\pm=\left(\pi_1\mp i\pi_2\right)/\sqrt 2$, and $\tau_\pm=\left(\tau_1\pm i\tau_2\right)/\sqrt 2$ are the raising and lowering operators in flavor space.

The order parameters of spontaneous chiral symmetry breaking and isospin symmetry breaking are respectively the expected values of the auxiliary fields $\langle\sigma\rangle$ and $\langle\pi_\pm\rangle$ ($\langle\pi_0\rangle=0$). Taking them to be real and constants, considering the relation between the chiral order parameter and the dynamical quark mass $\langle\sigma\rangle=m-m_0$, and denoting $\langle\pi_\pm\rangle$ by $\Delta$, the thermodynamic potential of the quark system in mean field approximation can be expressed in Euclidean space as
\begin{eqnarray}
\Omega &=& {\left(m-m_0\right)^2+\Delta^2\over 4G}\\
&&-{1\over\beta V}\text {Tr}\ln\left({\slashed p_E}-m-i\gamma_5\tau_1\Delta+{\mu_I\over2}\gamma_0\tau_3+\mu_5\gamma_0\gamma_5\right)\nonumber
\end{eqnarray}
with ${\slashed p_E}=i\gamma_0p_0-\gamma_ip_i$ and $\beta=1/T$, where the trace is taken over the quark spin, flavor, color, coordinate and momentum.
At zero temperature, the potential is explicitly expressed as
\begin{equation}
\Omega={\left(m-m_0\right)^2+\Delta^2\over 4G}-N_c\sum_{i,j=\pm}\int {d^3{\bf p}\over(2\pi)^3}E_{i,j}\ ,
\label{omega}
\end{equation}
where $E_{i,j}(p)=\sqrt{\xi_{i,j}^2(p)+\Delta^2}$ are the quasiparticle energies with $\xi_{i,j}(p)=\sqrt{(p+i\mu_5)^2+m^2}+j\mu_I/2$, and $N_c=3$ is the color degrees of freedom. The two order parameters $\sigma$ or $m$ and $\Delta$ are determined by the minimum of the thermodynamic potential, $\partial\Omega/\partial m=0$ and $\partial\Omega/\partial \Delta=0$, which lead to the two coupled gap equations,
\begin{eqnarray}
\label{gap1}
{(m-m_0)\over 2G}&=&N_c\sum_{i,j=\pm}\int{d^3{\bf p}\over(2\pi)^3}{\xi_{i,j}\over E_{i,j}}{m\over\sqrt{(|{\bf p}|+i\mu_5)^2+m^2}},\nonumber\\
{\Delta\over 2G}&=&N_c\sum_{i,j=\pm}\int{d^3{\bf p}\over(2\pi)^3}{\Delta\over E_{i,j}}\ .
\end{eqnarray}

We now discuss the effect of chiral imbalance on chiral symmetry restoration and pion superfluidity. Since the NJL model is not renormalizable, it is necessary to introduce a momentum cutoff $\Lambda$ to regulate the integrations in the gap equations (\ref{gap1}). Let us first consider the possible phases of the system in chiral limit with vanishing current quark mass $m_0=0$. In this case, there are only two parameters $G$ and $\Lambda$ in the model, they are fixed to be $G=5.01$ GeV$^{-2}$ and $\Lambda=0.65$ GeV by fitting the chiral condensate and pion decay constant in vacuum with $T=\mu_I=\mu_5=0$ \cite{zhuang}. As can be seen from the gap equations (\ref{gap1}), there always exists a trivial solution $m=0$ and $\Delta=0$, corresponding to the phase with chiral symmetry and without pion condensation. The nontrivial solutions are usually preferred by the system, as they correspond to lower thermodynamic potentials. However, the two gap equations with $m, \Delta\neq 0$ contradict with each other \cite{he}, and there are only two possible nontrivial solutions: the phase with spontaneous chiral symmetry breaking characterized by $m\neq 0$ and $\Delta=0$ and the phase with spontaneous isospin symmetry breaking described by $\Delta\neq 0$ and $m=0$. At finite isospin chemical potential, the pion superfluidity phase with $\Delta\neq 0$ is always the ground state of the system \cite{he}.

In vacuum, the critical coupling constant to keep spontaneous chiral symmetry breaking is $G_c=\pi^2/(2N_c\Lambda^2)$. By scaling the coupling constant $G$ by $G_c$, the gap equation for the nonzero pion condensate is reduced to
\begin{equation}
{G_c\over G}={1\over2}\sum_{i,j=\pm}\int_0^1 dp p^2{1\over E_{i,j}}\ ,
\label{regap}
\end{equation}
where $p, \mu_I, \mu_5$ and $\Delta$ are all scaled by the cutoff $\Lambda$.

\begin{figure}[!htb]
\centering
\includegraphics[width=0.4\textwidth]{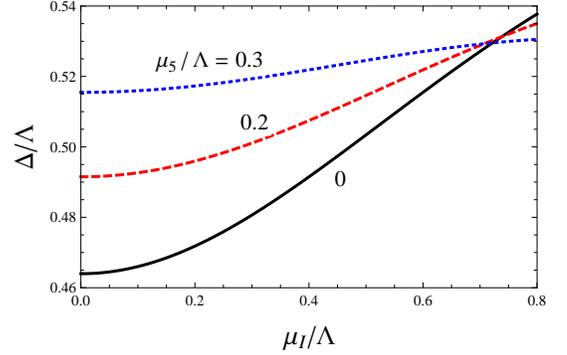}
\caption{(color online) The pion condensate $\Delta$ as a function of isospin chemical potential $\mu_I$ at fixed chiral chemical potential $\mu_5$ in chiral limit. All the quantities are scaled by the momentum cutoff $\Lambda$.}
\label{fig1}
\end{figure}
\begin{figure}[!htb]
\centering
\includegraphics[width=0.4\textwidth]{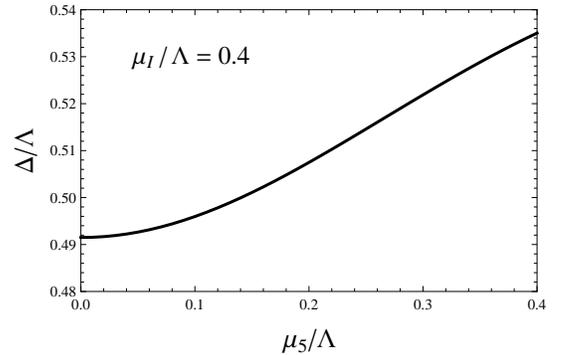}
\caption{The pion condensate $\Delta$ as a function of chiral chemical potential $\mu_5$ at fixed isospin chemical potential $\mu_I$ in chiral limit. All the quantities are scaled by the momentum cutoff $\Lambda$.}
\label{fig2}
\end{figure}
The scaled pion condensate is shown in Fig.\ref{fig1} as a function of scaled isospin chemical potential. As a first order phase transition, the pion condensate $\Delta$ jumps up from zero to a finite value at the critical isospin chemical potential $\mu_I^c=0$. While the condensate increases with increasing $\mu_I$ or $\mu_5$ in general case, it drops down with increasing $\mu_5$ when $\mu_I$ is large enough, see the up-right corner of Fig.\ref{fig1}. Considering the fact that $\mu_I$ in this case is already close to or even beyond the cutoff $\Lambda$, this dropping down is probably an artifact of the model. The $\mu_5$ dependence of the condensate at fixed $\mu_I/\Lambda=0.4$ is shown in Fig.\ref{fig2}, which displays a monotonous increase. Note that, while the pion condensate increases with chiral imbalance at reasonable isospin density, the critical point of pion superfluid is not affected by the chiral imbalance, it is always located at $\mu_I=0$.

\begin{figure}[!htb]
\centering
\includegraphics[width=0.4\textwidth]{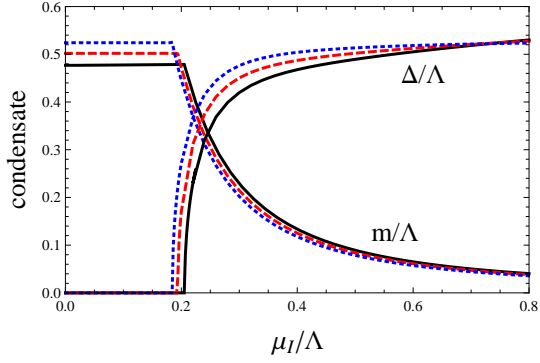}
\caption{(color online) The chiral and pion condensates $m$ and $\Delta$ as functions of isospin chemical potential $\mu_I$ in real case. All the quantities are scaled by the momentum cutoff $\Lambda$, and the solid, dashed and dotted lines correspond to fixed chiral chemical potential $\mu_5/\Lambda=0,\ 0.2$ and $0.3$. }
\label{fig3}
\end{figure}
\begin{figure}[!htb]
\centering
\includegraphics[width=0.41\textwidth]{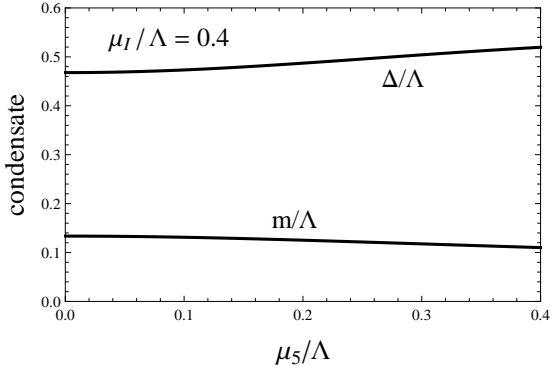}
\caption{The chiral and pion condensates $m$ and $\Delta$ as functions of chiral chemical potential $\mu_5$ at fixed isospin chemical potential $\mu_I$ in real case. All the quantities are scaled by the momentum cutoff $\Lambda$. }
\label{fig4}
\end{figure}

In real case with finite current quark mass, the three parameters of the model are fixed to be $G=4.93$ GeV$^{-2}$, $\Lambda=0.653$ GeV and $m_0=5$ MeV by fitting the pion mass $m_\pi=134$ MeV, pion decay constant $f_\pi=93$ MeV and quark condensate $\langle\sigma\rangle=-2\times (0.25\text {GeV})^3$ \cite{zhuang}. The isospin chemical potential dependence of the two order parameters is shown in Fig.\ref{fig3}. At $\mu_5/\Lambda=0$, the superfluid starts at the critical isospin chemical potential $\mu_I^c=m_\pi$ ($\mu_I^c/\Lambda=0.21$) \cite{he}. With increasing $\mu_5$, the critical point shifts towards the left but the pion condensate increases, which indicate a catalysis effect. The quark mass keeps as a constant in the normal phase at $\mu_I < \mu_I^c$ and drops down monotonously in the pion superfluid. With increasing $\mu_5$, the quark mass increases in the normal phase but decreases in the superfluid. Suppose the chiral phase transition happens at the same critical point as the pion superfluidity, the $\mu_5$ dependence of the quark mass shows an inverse catalysis effect on the chiral phase transition. This is similar to the lattice simulated magnetic field effect at finite temperature \cite{bali,bali2,bali3,bali4,bali5}: The chiral condensate increases at low temperature but the critical temperature is reduced. The chiral imbalance effect on the two order parameters at fixed isospin chemical potential is clearly shown in Fig.\ref{fig4}.

\section{pion superfluidity with magnetic field}
\label{section3}

Now we turn to the pion superfluidity in an external magnetic field. The Lagrangian density of the NJL model is written as
\begin{equation}
\label{njl2}
{\cal L}=\bar \psi\left(i\slashed D-m_0+{\mu_I\over 2}\gamma_0\tau_3\right)\psi+G\left[\left(\bar\psi\psi\right)^2+\left(\bar\psi i\gamma_5\boldsymbol\tau\psi\right)^2\right],
\end{equation}
where $D_\mu=\partial_\mu+iq A_\mu$ is the covariant derivative in flavor space with electric charges $q_u=2e/3$ and $q_d=-e/3$ for $u$ and $d$ quarks, and the potential $A_\mu=(0,0,Bx_1,0)$ defines a constant magnetic field along the $x_3-$axis through ${\bf B}=\nabla\times{\bf A}$. Since it is cumbersome to study pion superfluidity in an external magnetic field in quark models, as mentioned in the introduction, we derive here the phase transition line of pion superfluid by using the Ginzburg-Landau (GL) approach. Following the same procedure as in Section \ref{section2}, the thermodynamic potential in Minkowski space is expressed as
\begin{eqnarray}
\Omega={(m-m_0)^2+\Delta^2\over 4G}+{i\over V_4}\text{Tr}\ln\left(\begin{array}{cc}
(iG_u)^{-1}&-i\gamma_5\Delta\\
-i\gamma_5\Delta^*&(iG_d)^{-1}
\end{array}\right),
\end{eqnarray}
where $V_4$ is the space-time volume, and $(iG_f)^{-1}=i\slashed D_f-m\pm (\mu_I/2)\gamma_0\ (f=u,\ d)$ is the inverse quark propagator at mean field level. We now take Taylor expansion of $\Omega$ in terms of the pion condensate $\Delta$ around the critical point and keep only the first two terms,
\begin{eqnarray}
\Omega &=& {\Delta^2\over 4G}-{i\over 2V_4}\text{Tr}G_{xy}G_{yx}-{i\over 4V_4}\text {Tr}G_{xy_1}G_{y_1y_2}G_{y_2y_3}G_{y_3x}\nonumber\\
       &=& {\cal A}\Delta^2+{{\cal B}\over 2}\Delta^4
\end{eqnarray}
with the quark propagator in coordinate space
\begin{eqnarray}
G_{xy}=\left(\begin{array}{cc}
iG_u(x,y)&0\\
0&iG_d(x,y)
\end{array}\right)\left(\begin{array}{cc}
0&-i\gamma_5\Delta_y\\
-i\gamma_5\Delta^*_y&0
\end{array}\right)
\end{eqnarray}
and the coefficients ${\cal A}$ and ${\cal B}$ in the GL approximation
\begin{widetext}
\begin{eqnarray}
{\cal A}&=&{1\over 4G}+{i\over V_4}\text {Tr}\left[G_u(x,y)i\gamma_5G_d(y,x)i\gamma_5e^{-ie\int_x^y A^\mu dx_\mu}\right],\nonumber\\
{\cal B}&=&-{i\over V_4}\text {Tr}\Big[G_u(x,y_1)i\gamma_5G_d(y_1,y_2)i\gamma_5 G_u(y_2,y_3)i\gamma_5G_d(y_3,x)i\gamma_5e^{-ie\left(\int_{y_2}^{y_1}+\int_x^{y_3}\right)A^\mu dx_\mu}\Big].
\label{ab}
\end{eqnarray}

Note that in the expression of the coefficients, we have taken into account the interaction between the charged pion condensate and the magnetic field which is not included in the NJL model (\ref{njl2}). It is introduced by a straight-line link (the exponential in (\ref{ab})) between two points $\Delta_y$ and $\Delta^*_x$, and each condensate is linked only once. The quark propagators $G_f(x,y)$ can be evaluated with Schwinger approach \cite{schwinger},
\begin{eqnarray}
G_f(x,y) &=& e^{-iq_f\int_y^x\bar A_f^\mu dx_\mu}S_f(x-y),\\
S_f(x) &=& -i\int_0^\infty{ds\over 16(\pi s)^2}e^{-i\left[sm^2+{1\over 4s}\left(x_0^2-x_3^2-{\bf x}_\bot^2 B_f^s \cot B_f^s\right)\right]}
 B_f^s\left[\cot B_f^s-\gamma_1\gamma_2\right]\left[m+{1\over 2s}\left(\slashed x_0-\slashed x_3-B_f^s\left(\left(\slashed x_1+\slashed x_2\right)\cot B_f^s+\slashed x_{21}-\slashed
   x_{12}\right)\right)\right]\nonumber
\end{eqnarray}
with $B_f^s=q_fsB$, $\slashed x_\mu=\gamma_\mu x_\mu$, $\slashed x_{\mu\nu}=\gamma_\mu x_\nu$ and ${\bf x}_\bot^2=x_1^2+x_2^2$, where $\bar A_f^\mu=(\mp\mu_I/(2q_f),0,0,0)+A^\mu$ are effective potentials for $u$ and $d$ quarks, and the integration in the exponential from $y$ to $x$ is along a straight line. With the known quark propagators, the coefficients can be expressed in terms of the functions $S_f$,
\begin{eqnarray}
&& {\cal A} = {1\over 4G}+i{N_c\over V_4^2}\int d^4(x-y)\text {Tr}\left[e^{i{\mu_I\over 2}(x_0-y_0)}S_u(x-y)i\gamma_5e^{-i{\mu_I\over 2}(y_0-x_0)}S_d(y-x)i\gamma_5\right],\nonumber\\
&& {\cal B} = -i{N_c\over V_4^4}\int d^4(x-y_1)d^4(y_1-y_2)d^4(y_2-y_3)\times\\
&& \ \ \ \ \ \ \
   \text {Tr}\left[e^{i{\mu_I\over 2}(x_0-y_{10})}S_u(x-y_1)i\gamma_5e^{-i{\mu_I\over 2}(y_{10}-y_{20})}S_d(y_1-y_2)i\gamma_5e^{i{\mu_I\over 2}(y_{20}-y_{30})}S_u(y_2-y_3)i\gamma_5
   e^{-i{\mu_I\over 2}(y_{30}-x_0)}S_d(y_3-x)i\gamma_5\right],\nonumber
\end{eqnarray}
where the Wilson lines in the quark propagators are exactly canceled by the interaction between the charged pion condensate and the magnetic field, and therefore the gauge invariance is guaranteed. For convenience we transfer from the coordinate space to the energy-momentum space and switch on the temperature. The quark propagators in Euclidean space are then given as \cite{gusynin}
\begin{eqnarray}
S_f^E(\omega_n,{\bf k}) = -i\int_0^\infty ds e^{-s\left(m^2+\omega_n^2+k_3^2+{\bf k}_\bot^2 \tanh B_f^s/B_f^s\right)}\left(-\slashed k+m-i(k_{12}-k_{21})\tanh B_f^s\right)\left(1-i\gamma_1\gamma_2\tanh B_f^s\right)
\end{eqnarray}
with the Matsubara frequency $\omega_n=(2n+1)\pi T\ (n\in Z)$ of fermions. From the GL theory, the condition ${\cal A}=0$ determines the second order phase transition line of the pion superfluidity. After a straightforward calculation, we obtain
\begin{eqnarray}
{\cal A} &=&{1\over 4G}-N_c T\sum_n\int{d^3{\bf k}\over(2\pi)^3}\text {Tr}\left[S_u^E\left(\omega_n+i{\mu_I\over 2},{\bf k}\right)i\gamma_5S_d^E\left(\omega_n-i{\mu_I\over 2},{\bf k}\right)i\gamma_5\right]\\
&=&{1\over 4G}-4N_cT\sum_n\int {d^3{\bf k}\over(2\pi)^3}\int ds\int dt {\cal R}(s,t,\omega_n,{\bf k})\left\{\left(m^2+\omega_n^2+\left({\mu_I\over 2}\right)^2+k_3^2\right)\left(1+f_u(s)f_d(t)\right)+{\bf k}_\bot^2\left(1-f_u^2(s)\right)\left(1-f_d^2(t)\right)\right\}\nonumber
\end{eqnarray}
with $f_u(s)=\tanh B_u^s$, $f_d(t)=\tanh B_d^t$, transverse momentum ${\bf k}_\bot^2=k_1^2+k_2^2$ and
\begin{equation}
 {\cal R}(s,t,\omega_n,{\bf k})=e^{-s\left[m^2+\left(\omega_n+i\mu_I/2\right)^2+k_3^2+{\bf k}_\bot^2f_u(s)/B_u^s\right]-t\left[m^2+\left(\omega_n-i\mu_I/2\right)^2+k_3^2+{\bf k}_\bot^2f_d(t)/B_d^t\right]},
 \end{equation}
where the integrations over $s$, $t$ and ${\bf k}$ are divergent and a regularization scheme is needed. We take the way given in Refs.~\cite{schwinger,cao} and absorb all the divergence into the vacuum term which is then regularized by the three-momentum cutoff $\Lambda$. In this way the convergent coefficient ${\cal A}$ includes a vacuum part ${\cal A}_0$ and two magnetic field dependent parts ${\cal A}_{B1}$ and ${\cal A}_{B2}$,
\begin{eqnarray}
{\cal A}&=&{1\over4G}+{\cal A}_0+{\cal A}_{\rm B1}+{\cal A}_{\rm B2},\nonumber\\
{\cal A}_0 &=&-{N_c\Lambda^2\over 2\pi^2}\Bigg[\sqrt{1+\left({m\over \Lambda}\right)^2}-\left(\left({m\over \Lambda}\right)^2-{1\over 2}\left({\mu_I\over \Lambda}\right)^2\right)\ln\left({\Lambda\over m}
+\sqrt{1+{\Lambda^2\over m^2}}\right)\nonumber\\
&&-{\mu_I\over \Lambda}\sqrt{\left({m\over \Lambda}\right)^2-\left({\mu_I\over 2\Lambda}\right)^2}\tan^{-1}{{\mu_I\over \Lambda}\over 2\sqrt{\left(1+\left({m\over \Lambda}\right)^2\right)\left(\left({m\over \Lambda}\right)^2-\left({\mu_I\over 2\Lambda}\right)^2\right)}}\Bigg]+{N_c\over 2\pi^2}\sum_{i=\pm}\int_0^\infty k^2 dk{1\over E_i(k)}{2\over 1+e^{E_i(k)/T}},\nonumber\\
{\cal A}_{B1}&=&-{N_cT\over 4\pi^{3/2}}\sum_n\int_0^\infty {ds\over\sqrt s}\int_{-1}^1 dv e^{-s\left(m^2+\left(\omega_n+iv{\mu_I\over 2}\right)^2-\left({\mu_I\over 2}\right)^2(1-v^2)\right)}\Bigg[
{1\over s}\left({1\over \left(f_u\left({1+v\over2}s\right)/B_u^s+f_d\left({1-v\over 2}s\right)/B_d^s\right)^2}-1\right)\nonumber\\
&&+\left(m^2+\omega_n^2+{\mu}^2+{1\over 2s}\right)\left({1\over f_1\left({1+v\over2}s\right)/B_1^s+f_2\left({1-v\over 2}s\right)/B_2^s}-1\right)\Bigg],\nonumber\\
{\cal A}_{B2}&=&-{N_cT\over 4\pi^{3/2}}\sum_n\int_0^\infty {ds\over\sqrt s}\int_{-1}^1 dv e^{-s\left(m^2+\left(\omega_n+iv{\mu_I\over 2}\right)^2-\left({\mu_I\over 2}\right)^2(1-v^2)\right)}\Bigg[
{1\over s}{f_u^2\left({1+v\over 2}s\right)f_d^2\left({1-v\over 2}s\right)-f_u^2\left({1+v\over 2}s\right)-f_d^2\left({1-v\over 2}s\right)\over \left(f_u\left({1+v\over 2}s\right)/B_u^s+f_d\left({1-v\over 2}s\right)/B_d^s\right)^2}\nonumber\\
&&+{\left(m^2+\omega_n^2+\left({\mu_I\over 2}\right)^2+{1\over 2s}\right)f_u\left({1+v\over 2}s\right)f_d\left({1-v\over 2}s\right)\over f_u\left({1+v\over2}s\right)/B_u^s+f_d\left({1-v\over 2}s\right)/B_d^t}\Bigg]
\end{eqnarray}
with $E_i(k)=\sqrt{k^2+m^2}+i\mu_I/2$. It should be remarked that there is an implicit condition $m>\mu_I/2$ here, which makes the integrations over $s$ convergent in ultraviolet domain and which is guaranteed at small pion condensate where the dynamical quark mass $m$ is relatively large. This condition is only due to Schwinger approach itself for introducing proper-time integral and there is no such restriction with Ritus method~\cite{Ritus}.

The derivation of the coefficient ${\cal B}$ is more tedious. After a careful calculation, we finally obtain its explicit expression,
\begin{eqnarray}
{\cal B} &=& 4N_cT\sum_n\int{d^3{\bf k}\over(2\pi)^3}\int ds dt ds' dt' {\cal R}(s,t,\omega_n,{\bf k}){\cal R}(s',t',\omega_n,{\bf k})\times\\
&&\Bigg\{{1\over 2}\left[\left(\omega_n^u\omega_n^d+k_3^2+m^2\right)^2+\left(m\mu_I\right)^2+\left(k_3\mu_I\right)^2\right]\sum_{l=\pm}
  \left(1+lf_u(s)\right)\left(1+lf_u(s')\right)\left(1+lf_d(t)\right)\left(1+lf_d(t')\right)\nonumber\\
&&+{\bf k}_\bot^4\left(1-f_u^2(s)\right)\left(1-f_u^2(s')\right)\left(1-f_d^2(t)\right)\left(1-f_d^2(t')\right)
+{\bf k}_\bot^2\Big[4\left(\omega_n^u\omega_n^d+k_3^2+m^2\right)\left(1+f_u(s)f_d(t)\right)\left(1-f_u^2(s')\right)\left(1-f_d^2(t')\right)\nonumber\\
&&-\left(\left(\omega_n^u\right)^2+k_3^2+m^2\right)\left(1-f_u(s)f_u(s')\right)\left(1-f_d^2(t)\right)\left(1-f_d^2(t')\right)
-\left(\left(\omega_n^d\right)^2+k_3^2+m^2\right)\left(1-f_d(t)f_d(t')\right)\left(1-f_u^2(s)\right)\left(1-f_u^2(s')\right)\Big]\Bigg\}\nonumber
\end{eqnarray}
with the effective frequencies $\omega_n^u=\omega_n+i\mu_I/2$ and $\omega_n^d=\omega_n-i\mu_I/2$. It is not necessary to numerically calculate the coefficient ${\cal B}$, if we only focus on the phase transition line of the pion superfluid. What we are interested in here is its sign which determines the order of the phase transition and  thus the validity of the GL approach. Keeping in mind the symmetry between $s (t)$ and $s' (t')$ in the integrations, we have the inequality
\begin{eqnarray}
&&4\left(1+f_u(s)f_d(t)\right)\left(1-f_u^2(s')\right)\left(1-f_d^2(t')\right)-\left(1-f_u(s)f_u(s')\right)\left(1-f_d^2(t)\right)\left(1-f_d^2(t')\right)
-\left(1-f_d(t)f_d(t')\right)\left(1-f_u^2(s')\right)\left(1-f_u^2(s)\right)\nonumber\\
&\ge& \left(1-f_u^2(s')\right)\left(1-f_d^2(t')\right)
\left[4\left(1+f_u(s)f_d(t)\right)-\left(1+f_u^2(s)\right)\left(1-f_d^2(t)\right)-\left(1+f_d^2(t)\right)\left(1-f_u^2(s)\right)\right]\nonumber\\
&=&2\left(1+f_u(s)f_d(t)\right)^2\left(1-f_u^2(s')\right)\left(1-f_d^2(t')\right)\ge 0.
\end{eqnarray}
Then using the condition $m>\mu_I/2$ for small condensate $\Delta$ and the fact of small ratio Im ${\cal R}$/Re ${\cal R}$, the coefficient ${\cal B}$ is found to be positive definite around the transition line of the pion superfluid. Thus, the  transition from normal phase to pion superfluidity with non-vanishing magnetic field is proved to be of second order.

The phase transition line of the pion superfluid ${\cal A}=0$ is coupled with the dynamical quark mass $m$ which should be evaluated consistently with the gap equation for the chiral phase transition. Following Schwinger approach and taking the vacuum regularization scheme adopted above, we have the gap equation
\begin{eqnarray}
\label{gapm}
0&=& {m-m_0\over 2G}-{1\over \beta V}\sum_{f=u,d}\text {Tr}G_f(x,y)\nonumber\\
 &=& {m-m_0\over 2G}-{N_c m^2\over\pi^2}\left[\Lambda\sqrt{1+{\Lambda^2\over m^2}}-m\ln\left({\Lambda\over m}
+\sqrt{1+{\Lambda^2\over m^2}}\right)\right]+{N_cm\over\pi^2}\sum_{s=\pm}\int_0^\infty k^2 dk {1\over\sqrt{k^2+m^2}}{2\over 1+e^{E_s(k)/T}}\nonumber\\
&&-{N_cm\over 4\pi^2}\int_0^\infty{ds\over s^2} e^{-sm^2}\left[\vartheta_3\left({\pi\over 2}+i{\mu_I\over 4T},e^{-{1\over 4sT^2}}\right)\left({q_usB\over f_u(s)}-1\right)+\vartheta_3\left({\pi\over 2}-i{\mu_I\over 4T},e^{-{1\over 4sT^2}}\right)\left({q_dsB\over f_d(s)}-1\right)\right],
\end{eqnarray}
\end{widetext}
where $\vartheta_3(z,q)$ is the third Jacobi theta function obtained by working out the summation over the Matsubara frequency.

\begin{figure}[!htb]
\centering
\includegraphics[width=0.4\textwidth]{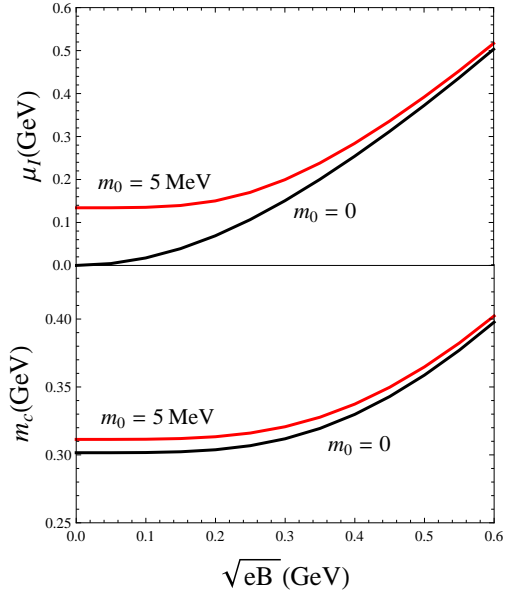}
\caption{(color online) The phase diagrams of pion superfluid in $\mu_I-B$ plane (upper panel) and the quark mass $m_c$ on the phase transition line (lower panel) in chiral limit ($m_0=0$) and real case ($m_0=5$ MeV). }
\label{fig5}
\end{figure}
We now show the phase diagram of pion superfluidity in an external magnetic field. We adopt the same set of parameters of the model as given in Section \ref{section2}. The phase diagram in $\mu_I-B$ plane at zero temperature is shown in the upper panel of Fig.\ref{fig5}. The two solid lines are the phase transition lines of pion superfluidity in chiral limit with $m_0=0$ and real case with $m_0=5$ MeV. The normal phase without pion condensate is under the corresponding line, and the pion superfluidity phase is above the line. In both cases, the critical isospin chemical potential $\mu_I^c$ increases with the strength of the magnetic field $B$, which indicates clearly the effect of inverse magnetic catalysis on the pion superfluidity. This is consistent with the lattice QCD result \cite{endrodi}. Note that the above analytic and numerical results depend on the condition $m>\mu_I/2$. To check if this condition is satisfied, we show in the lower panel of Fig.\ref{fig5} the dynamic quark mass $m_c$ as a function of the magnetic field on the phase transition line. By comparing it with the upper panel, there is always $m_c>\mu_I^c/2$.

\begin{figure}[!htb]
\centering
\includegraphics[width=0.4\textwidth]{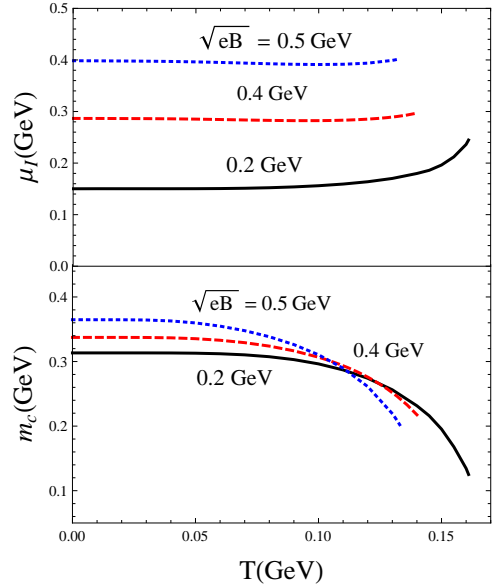}
\caption{(color online) The phase diagrams of pion superfluid in $\mu_I-T$ plane (upper panel) and the quark mass $m_c$ on the phase transition line (lower panel) at fixed magnetic field and in real case with $m_0$=5 MeV.}
\label{fig6}
\end{figure}
The phase diagram in $\mu_I-T$ plane is shown in the upper plane of Fig.\ref{fig6} in real case with $m_0$=5 MeV. The phase transition line at a fixed magnetic field separates the normal phase at low $\mu_I$ and the pion superfluidity at high $\mu_I$. The increasing critical isospin chemical potential $\mu_I^c$ with the magnetic field indicates again the inverse catalysis effect on the pion superfluidity. While the temperature dependence of $\mu_I^c$ is very smooth at low temperature, it goes up quickly at high temperature. The quark mass on the phase transition line is shown in the lower panel of Fig.\ref{fig6}. It decreases with temperature due to the gradual chiral symmetry restoration. The small triangle structure in the intersection region shows the de Haas--van Alphan oscillation~\cite{haas,landau} induced by the interplay among the field, chemical potential and temperature. The reason why the de Haas--van Alphan oscillation doesn't show up at low temperature is that the gap equation (\ref{gapm}) only depends on $\mu_I$ weakly now as the third Jacobi theta function $\vartheta_3(z,q)\approx1$ in the forth term on the right hand side.

\section{Color superconductivity with chiral imbalance and magnetic field}
\label{section4}

We discuss now color superconductivity including simultaneously the chiral imbalance and magnetic field effects. The Lagrangian density of the extended NJL model with baryon chemical potential $\mu_B$, chiral chemical potential $\mu_5$ and external magnetic field $B$ is defined as
\begin{eqnarray}
{\cal L}&=&\bar{\psi}\left[i\gamma^\mu\left(D_\mu-igT_8G_\mu^8\right)-m_0+\mu_B\gamma_0+\mu_5\gamma_0\gamma_5\right]\psi\\
&&+G_S\left[\left(\bar\psi\psi\right)^2+\left(\bar\psi i\gamma_5\boldsymbol{\tau}\psi\right)^2\right]
+G_D\left(i\bar\psi_C\varepsilon\epsilon_3\gamma_5\psi\right)\left(i\bar\psi\varepsilon\epsilon_3\gamma_5\psi_C\right),\nonumber
\end{eqnarray}
where $\psi_C=C\bar\psi^T$ and $\bar\psi_C=\psi^TC$ are charge-conjugate spinors with $C=i\gamma_2\gamma_0$, $\varepsilon_{ij}$ and $(\epsilon_3)_{ab}=\epsilon_{ab3}$ are respectively the antisymmetric matrices in flavor and color spaces with indexes $i,j=(u,d)$ and $a,b=(r,g,b)$, the term $-\bar\psi ig\lambda_8G_\mu^8\psi/2$ accounts for the interaction between quarks and massless gluons \cite{alford, litim} with $\lambda_8$ being the 8th Gell-Mann matrix, and the coupling constants $G_S$ and $G_D$ in the scalar and diquark channels are related to each other by the Fierz transformation $G_D=3G_S/4$ \cite{huang}.

Using the "rotated" charge operator $\tilde Q=Q\otimes \boldsymbol {1_c}-\boldsymbol {1_f}\otimes \lambda_8/(2\sqrt 3)$, a massless $U_{em}(1)$ field $\tilde A_\mu=A_\mu\cos\theta+G_\mu^8\sin\theta$ and a massive gluon field $\tilde G_\mu^8=G_\mu^8\cos\theta-A_\mu\sin\theta$ are obtained \cite{fayazbakhsh}. By neglecting the massive gluon field and replacing the massless field with an external magnetic field $A_\mu=(0,0,Bx_1,0)$ as $\theta$ is small, the following simplified Lagrangian is derived,
\begin{eqnarray}
{\cal L}&=&\bar\psi\left[i\gamma^\mu \tilde D_\mu-m_0+\mu_B\gamma_0+\mu_5\gamma_0\gamma_5\right]\psi\\
&&+G_S\left[\left(\bar\psi\psi\right)^2+\left(\bar\psi i\gamma_5\boldsymbol \tau\psi\right)^2\right]
+G_D\left(i\bar\psi_C\varepsilon\epsilon_3\gamma_5\psi\right)\left(i\bar\psi\varepsilon\epsilon_3\gamma_5\psi_C\right)\nonumber
\end{eqnarray}
with $\tilde{D}_\mu=\partial_\mu+i\tilde{Q}eA_\mu^e$ and $\tilde{Q}=diag(q_{\rm ur},q_{\rm ug},q_{\rm ub},q_{\rm dr},q_{\rm dg},q_{\rm db})$ $=diag(1/2,1/2,1,-1/2,-1/2,0)$ in the flavor and color spaces.

Introducing the auxiliary fields
\begin{eqnarray}
\sigma&=&-2G_S\bar\psi\psi,\nonumber\\
\boldsymbol\pi&=&-2G_S\bar\psi i\gamma_5\boldsymbol \tau\psi,\nonumber\\
\Delta_c&=&-2G_D i\bar\psi_C\varepsilon\epsilon_3\gamma_5\psi,\nonumber\\
\Delta_c^*&=&-2G_D i\bar\psi\varepsilon\epsilon_3\gamma_5\psi_C,
\end{eqnarray}
the Lagrangian density becomes
\begin{eqnarray}
{\cal L}&=&\bar\psi\left[i\gamma^\mu \tilde D_\mu-m_0-\sigma-i\gamma_5\boldsymbol\tau\cdot\boldsymbol \pi+\mu_B\gamma_0+\mu_5\gamma_0\gamma_5\right]\psi\\
&&-{1\over2}\left[\Delta_c i\bar\psi\varepsilon\epsilon_3\gamma_5\psi_C
+\Delta_c^*i\bar\psi_C\varepsilon\epsilon_3\gamma_5\psi\right]
-{\sigma^2+\boldsymbol\pi^2\over 4G_S}-{\Delta_c\Delta_c^*\over 4G_D}.\nonumber
\end{eqnarray}
In mean field approximation, supposing constant condensates $\langle\sigma\rangle=m-m_0$, $\langle\boldsymbol\pi\rangle=0$ and $\langle\Delta_c\rangle=\langle\Delta_c^*\rangle=\Delta_c$, we have the partition function
\begin{eqnarray}
Z&=&\int\left[d\bar\psi\right]\left[d\psi\right]\exp\bigg\{-i\int d^4x\bigg[{(m-m_0)^2\over 4G_S}+{\Delta_c^2\over 4G_D}\nonumber\\
&&-\bar\psi\left(i\gamma^\mu \tilde D_\mu-m+\mu_B\gamma_0+\mu_5\gamma_0\gamma_5\right)\psi\nonumber\\
&&+{1\over2}\Delta_c\left(i\bar\psi\varepsilon\epsilon_3\gamma_5\psi_C+
i\bar\psi_C\varepsilon\epsilon_3\gamma_5\psi\right)\bigg]\bigg\},
\end{eqnarray}
where the functional integrations over $\psi$ and $\bar\psi$ can be accomplished in the Nambu-Gorkov space. With the help of the charge projectors
\begin{equation}
P_q=\left\{\begin{array}{l}
diag(0,0,0,1,1,0)\ ,\qquad\qquad q=-1/2\\
diag(0,0,0,0,0,1)\ ,\qquad\qquad q=0\\
diag(1,1,0,0,0,0)\ ,\qquad\qquad q=1/2\\
diag(0,0,1,0,0,0)\ ,\qquad\qquad q=1,
\end{array}\right.
\end{equation}
the bispinors in Nambu-Gorkov space are defined as
\begin{equation}
\bar\Psi_q=\left(\begin{array}{cc}
\bar\psi_q & \bar\psi_{-q}^C
\end{array}\right),\;\Psi_q=\left(\begin{array}{c}
\psi_q\\ \psi_{-q}^C
\end{array}\right)
\end{equation}
with $\psi_q=P_q\psi$, and the partition function becomes
\begin{eqnarray}
Z&=&\int\left[d\bar\psi\right]\left[d\psi\right]\exp\bigg\{-i\int d^4x\bigg[{(m-m_0)^2\over 4G_S}+{\Delta_c^2\over 4G_D}\nonumber\\
&&-{1\over2}\sum_q \bar\Psi_qS_q\Psi_q\bigg]\bigg\},
\end{eqnarray}
where the inverse fermion propagators in Nambu-Gorkov space are defined as
\begin{eqnarray}
S_q=\left\{\begin{array}{l}\left(\begin{array}{cc}
i(D^+_q)^{-1}&0\\ 0&i(D^-_q)^{-1}
\end{array}\right)\qquad\qquad\:\:\; \ \ \ \ \ \ \ \ \ q=0,1\\
\\
\left(\begin{array}{cc}
i(D^+_q)^{-1}&-i\Delta_c\tau_2\lambda_2\gamma_5P_{-q}\\ -i\Delta_c\tau_2\lambda_2\gamma_5P_q & i(D^-_q)^{-1}\end{array}\right)\quad\:\: q=\pm{1\over2}
\end{array}\right.
\end{eqnarray}
with $i(D^\pm_q)^{-1}(x)=i\gamma^\mu \tilde D_\mu-m\pm \mu_B\gamma_0+\mu_5\gamma_0\gamma_5$ and the second Gell-mann matrix $\lambda_2$ in color space. After integrating out the bispinors and eliminating the double counting, the thermodynamic potential $\Omega=-T/V\ln Z$ can be expressed as
\begin{eqnarray}
\Omega&=&{(m-m_0)^2\over 4G_S}+{\Delta_c^2\over 4G_D}+{i\over2 V_4}\sum_q \text {Tr}\ln S_q.
\end{eqnarray}

For $q=0,1$, the trace can be worked out easily \cite{fukushima2},
\begin{eqnarray}
&&\Omega_0+\Omega_1\nonumber\\
&=&-\sum_{i,j=\pm}\bigg\{\int {d^3{\bf p}\over(2\pi)^3}\left[{E_i^0\over2}+T\ln\left(1+e^{-\left(E_i^0+j\mu_B\right)/T}\right)\right]\nonumber\\
&&+\sum_{n,p_3}\left[{E_i^n\over 2}+T\ln\left(1+e^{-\left(E_i^n+j\mu_B\right)/T}\right)\right]\bigg\},
\end{eqnarray}
where $\sum_{n,p_3}={eB\over 2\pi}\sum_{n=0}\alpha_n\int{dp_3\over 2\pi}$ with $\alpha_n=(2-\delta_{n0})/2$ is the summation over the Landau level and the integration over the longitudinal momentum in the external magnetic field,  and the quasiparticle energies are defined as
\begin{eqnarray}
E_i^0(p) &=&\sqrt{\left(p+i\mu_5\right)^2+m^2}\ ,\nonumber\\
E_i^n(p_3)&=&\sqrt{\left(\sqrt{2neB+p_3^2}+i\mu_5\right)^2+m^2}\ .
\label{EE1}
\end{eqnarray}

For $q=\pm 1/2$, the trace can be evaluated exactly, due to the same sign of the $\mu_5\gamma_0\gamma_5$ terms and the same electric charge in the diagonal terms of $S_q$. Following a similar procedure as in Ref.~\cite{fayazbakhsh2}, we get
\begin{equation}
\Omega_{+1/2}+\Omega_{-1/2}=-\sum_{i,j=\pm}\sum_{n,p_3}\left[E_{i,j}^n+2T\ln\left(1+e^{-E_{i,j}^n/T}\right)\right]
\end{equation}
with the quasiparticle energies
\begin{eqnarray}
E_{i,j}^n(p_3)=\sqrt{\left(E_i^{n/2}\left(p_3\right)+j\mu_B\right)^2+\Delta_c^2}\ .
\end{eqnarray}

For simplicity, we consider in the following only the case at zero temperature $T=0$ and in chiral limit with $m_0=0$. In this case, the thermodynamic potential can be simplified as
\begin{eqnarray}
\Omega &=& {m^2\over 4G_S}+{\Delta_c^2\over 4G_D}-\sum_{i,j=\pm}\Bigg\{\nonumber\\
&&\int {d^3{\bf p}\over (2\pi)^3}\left[{E_i^0\over 2}\theta(E_i^0-\mu_B)+{\mu_B\over 2}\theta(\mu_B-E_i^0)\right]\nonumber\\
&&+\sum_{n,p_3}\left[{E_i^n\over 2}\theta(E_i^n-\mu_B)+{\mu_B\over 2}\theta(\mu_B-E_i^n)+E_{i,j}^n\right]\Bigg\}
\end{eqnarray}
with the unit step function $\theta(x)$. To remove the divergence in the integrations, we use here the Pauli-Villars regularization scheme \cite{klevansky}, which is much softer than the hard three-momentum cutoff and works better in the case with a nonzero chiral imbalance, in comparison with the vacuum regularization used in Section \ref{section3}. From the numerical results shown below, the Pauli-Villars regularization will greatly suppress the artificial oscillation of the quark mass in external magnetic field \cite{fayazbakhsh}. In this way, the regularized version of the thermodynamic potential is given as
\begin{eqnarray}
\Omega &=&{m^2\over 4G_S}+{\Delta_c^2\over 4G_D}-\sum_{i,j=\pm}\sum_{l=0}^2 f_l\Bigg\{\nonumber\\
&&\int{d^3{\bf p}\over(2\pi)^3}\left[{{\cal E}_i^0\over 2}\theta({\cal E}_i^0-\mu_B)+{\mu_B\over2}\theta(\mu_B-{\cal E}_i^0)\right]\nonumber\\
&&+\sum_{n,p_3}\left[{{\cal E}_i^n\over 2}\theta({\cal E}_i^n-\mu_B)+{\mu_B\over 2}\theta(\mu_B-{\cal E}_i^n)+{\cal E}_{i,j}^n\right]\Bigg\}
\label{omega}
\end{eqnarray}
with $f_l=3l^2-6l+1$ and the regularized quasiparticle energies
 \begin{eqnarray}
{\cal E}_i^0(l,p)&=&\sqrt{\left(E_i^0(p)\right)^2+l\Lambda^2},\nonumber\\
{\cal E}_i^n(l,p_3)&=&\sqrt{\left(E_i^n(p_3)\right)^2+l\Lambda^2},\nonumber\\
{\cal E}_{i,j}^n(l,p_3)&=&\sqrt{\left({\cal E}_i^{n/2}\left(l,p_3\right)+j\mu_B\right)^2+\Delta_c^2}.
 \end{eqnarray}

The chemical potential and magnetic field dependence of the order parameters $m(\mu_B,\mu_5,B)$ and $\Delta_c(\mu_B,\mu_5,B)$ is controlled by the minimum of the thermodynamic potential $\partial\Omega/\partial m=0$ and $\partial\Omega/\partial \Delta_c=0$ which lead to the two gap equations
\begin{eqnarray}
{m\over2G_S}&=&m\sum_{i,j=\pm}\sum_{l=0}^2f_l\bigg\{\int{d^3{\bf p}\over (2\pi)^3}{1\over 2{\cal E}_i^0}\theta({\cal E}_i^0-\mu_B)\nonumber\\
&&+\sum_{n,p_3}\bigg[{1\over 2{\cal E}_i^n}\theta({\cal E}_i^n-\mu_B)+{{\cal E}_i^{n/2}+j\mu_B\over {\cal E}_i^{n/2}{\cal E}_{i,j}^n}\bigg]\bigg\},\nonumber\\
{\Delta_c\over 2G_D}&=&\Delta_c\sum_{i,j=\pm}\sum_{l=0}^2\sum_{n,p_3}{f_l\over {\cal E}_{i,j}^n}.
\label{gaps}
\end{eqnarray}
Note that with the Pauli-Villars regularization we used, the thermodynamic potential (\ref{omega}) is still divergent, but the gap equations (\ref{gaps}) we are interested in are convergent. In fact, the quantity $\Omega$ corresponds to the pressure except for a sign, and only the pressure relative to the vacuum can be measured. While $\Omega(\mu_B,\mu_5,B)$ is divergent, the physical potential $\Omega(\mu_B,\mu_5,B)-\Omega(0,0,0)$ is convergent.

It is easy to see that a trivial solution of the gap equations is $m=\Delta_c=0$, which corresponds to the phase with both chiral and color symmetries. The phase with both chiral and color symmetry breaking described by $m,\Delta_c\neq 0$ does not exist, due to the conflict between the two gap equations in chiral limit. The possible nontrivial solutions include the normal phase with $m\neq 0, \Delta_c=0$ and the color superconductor with $m=0, \Delta_c\neq 0$.

\begin{figure}[!htb]
\centering
\includegraphics[width=0.4\textwidth]{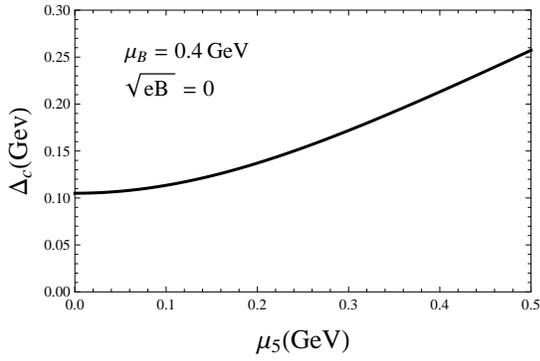}
\caption{The color condensate $\Delta_c$ as a function of chiral imbalance $\mu_5$ at fixed baryon chemical potential $\mu_B=0.4$ GeV and vanishing magnetic field $B=0$. }
\label{fig7}
\end{figure}

We choose the Pauli-Villars parameters as $\Lambda=0.859$ GeV, $G_S\Lambda^2=2.84$ and $G_D=3G_S/4$ \cite{itzykson,klevansky}. Let us first look at the chiral imbalance effect on color superconductivity, shown in Fig.\ref{fig7} at $\mu_B=0.4$ GeV and $B=0$. Since the diquark condensate can be expressed in terms of chiral quarks as
\begin{eqnarray}
\Delta_c&=&-2G_Di\langle\bar\psi_C\varepsilon\epsilon_3\gamma_5\psi\rangle\nonumber\\
&=&2G_D\left[\langle\psi_L^T\gamma_2\gamma_0\varepsilon\epsilon_3\gamma_5\psi_L\rangle+\langle\psi_R^T\gamma_2\gamma_0\varepsilon\epsilon_3\gamma_5\psi_R\rangle\right],
\end{eqnarray}
the two condensed quarks are with the same chirality, and therefore the chiral imbalance should enhance the color condensate.
\begin{figure}[!htb]
\centering
\includegraphics[width=0.4\textwidth]{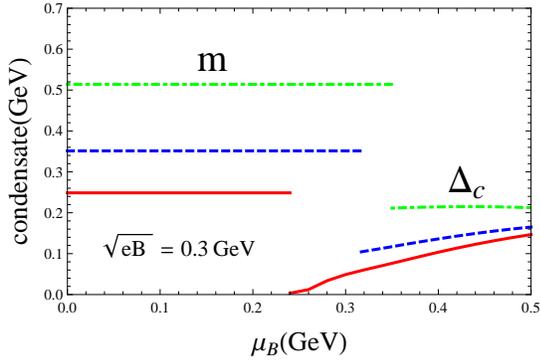}
\caption{(color online) The chiral and color condensates $m$ and $\Delta_c$ as functions of baryon chemical potential $\mu_B$ at fixed magnetic field $eB=0.3$ GeV and chiral imbalance $\mu_5=0$ (solid lines), $0.2$ GeV (dashed lines) and $0.4$ GeV (dot-dashed lines). }
\label{fig8}
\end{figure}
\begin{figure}[!htb]
\centering
\includegraphics[width=0.4\textwidth]{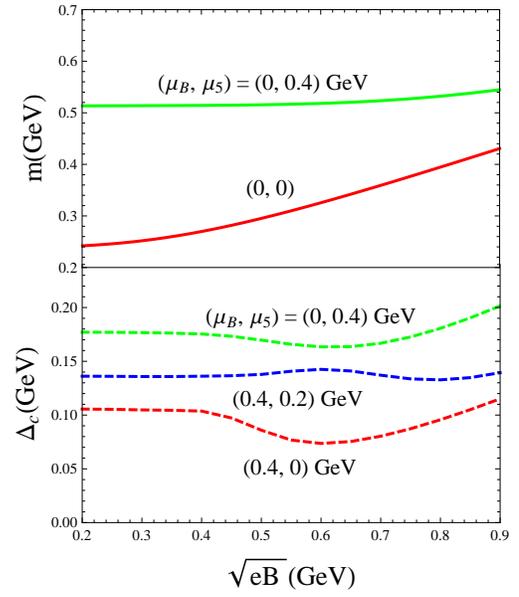}
\caption{(color online) The chiral and color condensates $m$ (upper panel) and $\Delta_c$ (lower panel) as functions of magnetic field $B$ at fixed baryon and chiral chemical potentials $\mu_B$ and $\mu_5$. }
\label{fig9}
\end{figure}
\begin{figure}[!htb]
\centering
\includegraphics[width=0.4\textwidth]{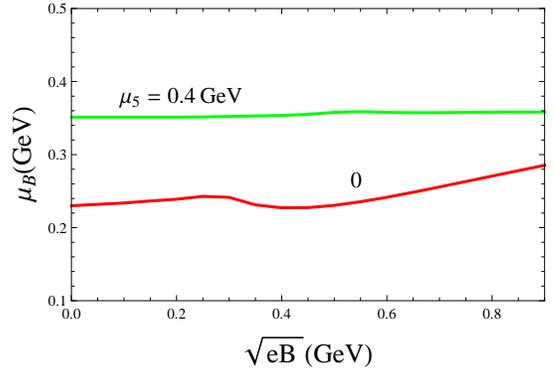}
\caption{(color online) The phase diagram of chiral restoration and color superconductivity in $\mu_B-B$ plane at fixed chiral chemical potential $\mu_5$.}
\label{fig10}
\end{figure}

The baryon chemical potential dependence of the two order parameters $m$ and $\Delta_c$ of chiral restoration and color superconductivity at fixed chiral imbalance and magnetic field is shown in Fig.\ref{fig8}. The system is in normal phase with chiral symmetry breaking ($m\neq 0$) at low $\mu_B$ and color superconductivity phase ($\Delta_c\neq 0$) at high $\mu_B$. The chiral condensate keeps as a constant in the normal phase, suddenly jumps down to zero at the critical baryon chemical potential $\mu_B^c$, and remains zero in the color superconductivity phase. In contrast, the color condensate remains zero in the normal phase, suddenly jumps up to a nonzero value at $\mu_B^c$, and then increases with $\mu_B$ in the color superconductivity phase. The two phase transitions at the critical point $\mu_B^c$ are both of first order. With increasing chiral imbalance, both the chiral and color condensates are enhanced, but the critical point shifts towards right. This indicates a stronger catalysis effect of chiral imbalance on the chiral condensate than on the color condensate.

The magnetic field effect is shown in Fig.\ref{fig9}. While the de Haas--van Alphan oscillation for quark mass $m$ is not observed, which is consistent with the results in Refs.~\cite{chao,cao,boomsma}, the oscillation shows up for color condensate $\Delta_c$ at any combination of chiral and baryon chemical potentials, similar to the results in Refs.~\cite{noronha,fukushima,fayazbakhsh}. We would like to point out the similarity between the two chemical potentials in the gap equation for color superconductivity. In this phase with $m=0$ and $\Delta_c\neq 0$, the excitation energy becomes
\begin{eqnarray}
E_{i,j}^n(p_3)=\sqrt{\left(\sqrt{neB+p_3^2}+i\mu_5+j\mu_B\right)^2+\Delta_c^2}\ ,
\end{eqnarray}
and we can introduce two combined chemical potentials $\mu_\pm=|\mu_5\pm\mu_B|$ instead of $\mu_B$ and $\mu_5$. In this way, there should be no difference between the case with $(\mu_B\neq 0, \mu_5=0)$ and the case with $(\mu_B=0, \mu_5\neq 0)$. The discrepancy between the line with $(\mu_B,\mu_5)=(0.4, 0)$ GeV and the line with $(\mu_B,\mu_5)=(0, 0.4)$ GeV shown in the lower panel of Fig.\ref{fig9} is due to the Pauli-Villars regularization which makes $\mu_5$ different from $\mu_B$ by introducing two large masses $\Lambda$ and $\sqrt{2}\Lambda$. However, the regularization mainly leads to a global shift between the two lines, it does not change the structure of the magnetic field dependence.

Finally, we show in Fig.\ref{fig10} the phase diagram of chiral restoration and color superconductivity in the plane of baryon chemical potential and magnetic field at fixed chiral imbalance. As can be seen, the de Haas--van Alphan oscillation for the critical baryon chemical potential $\mu_B^c$ shows up clearly at vanishing chiral imbalance. Such oscillation for chiral phase transition has been observed and well explained in Ref.~\cite{ebert}. At large chiral chemical potential $\mu_5=0.4$ GeV, the critical baryon chemical potential is almost a constant $\mu_B^c=0.35$ GeV, the oscillation is washed away by the strong chiral imbalance.

\section{conclusions}
\label{section5}

The phase transitions with chiral imbalance and magnetic field in strongly interacting quark matter are investigated at finite isospin and baryon densities. In the frame of extended NJL models, we focused on the magnetic effects on the pion superfluidity and color superconductivity and the related de Haas--van Alphan oscillations. For pion superfluidity, due to the problem of mixed quark propagators at different Landau levels, we treated the chiral imbalance and external magnetic field separately and take into account the interaction between the charged pion condensate and the magnetic field in the GL approach. For color superconductivity, we self-consistently studied the chiral imbalance and magnetic field effects on the order parameters of chiral restoration and color superconductivity.

For the pion superfluidity, the critical value of isospin chemical potential is suppressed by the chiral imbalance but enhanced by the external magnetic field, indicating respectively a catalysis and an inverse catalysis effect. The latter is consistent with the lattice simulated magnetic effect. For the color superconductivity, the chiral imbalance leads to a weaker catalysis effect compared to chiral symmetry breaking, and the magnetic field results in a de Haas--van Alphan oscillation on the phase transition line. However, the oscillation is washed away when the chiral imbalance is strong enough.

{\bf Acknowledgments:} The work is supported by the NSFC under grant No. 11335005 and the MOST under grant Nos. 2013CB922000 and 2014CB845400.


\begin{thebibliography}{99}
\bibitem{ando} T.Ando, Y.Matsumoto, and Y.Uemura, J. Phys. Soc. Jpn. {\bf 39}, 279(1975).
\bibitem{klitzing} K.Klitzing, G.Dorda, and M.Pepper, Phys. Rev. Lett. {\bf 45}, 494(1980).
\bibitem{tsui} D.C.Tsui, H.L.Stormer, and A.C.Gossard, Phys. Rev. Lett. {\bf 48}, 1559(1982).
\bibitem{kane} C.L.Kane and E.J.Mele, Phys. Rev. Lett. {\bf 95}, 226801(2005).
\bibitem{bernevig} B.A.Bernevig and S.C.Zhang, Phys. Rev. Lett. {\bf 96}, 106802(2006).
\bibitem{chang} C.Chang et al., Science {\bf 340}, 167(2013).
\bibitem{haas} W.J.de Haas and P.M.van Alphen, Proc. Acad. Sci. (Amsterdam), {\bf 33}, 1106(1930).
\bibitem{landau} L.D.Landau and E.M.Lifshitz, \emph{Statistical Physics}, Pergamon, New York, 1980.
\bibitem{ebert} D.Ebert, K.G.Klimenko, M.A.Vdovichenko, and A.S.Vshivtsev, Phys. Rev. {\bf D 61}, 025005(1999).
\bibitem{noronha} J.L.Noronha and I.A.Shovkovy, Phys. Rev. {\bf D 76}, 105030(2007).
\bibitem{fukushima} K.Fukushima, and H.J.Warringa, Phys. Rev. Lett. {\bf 100}, 032007(2008).
\bibitem{fayazbakhsh} Sh.Fayazbakhsh and N.Sadooghi, Phys. Rev. {\bf D 83}, 025026(2011).
\bibitem{bali} G.S.Bali, F.Bruckmann, G.Endrodi, Z.Fodor, S.D.Katz, S.Krieg, A.Schafer, and K.K.Szabo, JHEP {\bf 1202}, 44(2012).
\bibitem{bali2} G.S.Bali, F.Bruckmann, G.Endrodi, Z.Fodor, S.D.Katz, and A.Schafer, Phys. Rev. {\bf D 86} 071502(2012).
\bibitem{bali3} G.S.Bali, F.Bruckmann, M.Constantinou, M.Costa, G.Endrodi, S.D.Katz, H.Panagopoulos, and A.Schafer, Phys. Rev. {\bf D 86}, 094512(2012).
\bibitem{bali4} G.S.Bali, F.Bruckmann, G.Endrodi, F.Gruber, and A.Schaefer, JHEP {\bf 1304}, 130(2013).
\bibitem{bali5} G.S.Bali, F.Bruckmann, G.Endrodi, S.D.Katz, and A.Schaefer,  arXiv:1406.0269.
\bibitem{adler} S.L.Adler, Phys. Rev. {\bf 177}, 2246(1969).
\bibitem{bell} J.S.Bell and R.Jackiw, Nuovo Cim. {\bf A 60}, 47(1969).
\bibitem{kharzeev} D.Kharzeev, Phys. Lett. {\bf B 633}, 260(2006).
\bibitem{kharzeev2} D.Kharzeev, L.D.McLerran, and H.J.Warringa, Nucl. Phys. {\bf A 803}, 227(2008).
\bibitem{skokov} V.Skokov, A.Illarionov, and V.Toneev, Int. J. Mod. Phys. {\bf A 24}, 5925(2009).
\bibitem{son} D.T.Son and A.R.Zhitnitsky, Phys. Rev. {\bf D 70}, 074018(2004).
\bibitem{metlitski} M.A.Metlitski and A.R.Zhitnitsky, Phys. Rev. {\bf D 72}, 045011(2005).
\bibitem{xuguang}  X.G.Huang and J.Liao, Phys. Rev. Lett. {\bf 110}, 23, 232302(2013). 
\bibitem{fukushima2} K.Fukushima, M.Ruggieri, and R.Gatto, Phys. Rev. {\bf D 81}, 114031(2010).
\bibitem{chao} J.Chao, P.Chu, and M.Huang, Phys. Rev. {\bf D 88}, 054009(2013).
\bibitem{cao} G.Cao, L.He, and P.Zhuang, Phys. Rev. {\bf D 90}, 056005(2014).
\bibitem{kang} X.Kang, M.Jin, J.Xiong, and J.Li, arXiv:1310.3012.
\bibitem{loewe} M.Loewe, C.Villavicencio, and R.Zamora, Phys. Rev. {\bf D 89}, 016004(2014).
\bibitem{endrodi} G.Endrodi, Phys. Rev. {\bf D90}, 094501(2014).
\bibitem{zhuang} P.Zhuang, J.Hufner, and S.P.Klevansky, Nucl. Phys. {\bf A 576}, 525(1994).
\bibitem{he} L.He, M.Jin, and P.Zhuang, Phys. Rev. {\bf D 71}, 116001(2005).
\bibitem{schwinger} J.Schwinger, Phys. Rev. {\bf 82}, 664(1951).
\bibitem{gusynin} V.P.Gusynin, V.A.Miransky, and I.A.Shovkovy, Phys. Rev. Lett. {\bf 73}, 3499(1994); Phys. Rev. {\bf D 52}, 4718(1995); Phys. Lett. {\bf B 349}, 477(1995); Nucl. Phys. {\bf B 462}, 249(1996).
\bibitem{Ritus}        {V. I. Ritus, Ann. Phys. (Berlin) {\bf 69}, 555 (1972);
                        V. I. Ritus, Sov. Phys. JETP {\bf 48}, 788 (1978).}
\bibitem{alford} M.G.Alford, K.Rajagopal, and F.Wilczek, Nucl. Phys. {\bf B 537}, 443(1999).
\bibitem{litim} D.F.Litim and C.Manuel, Phys. Rev. {\bf D 64}, 094013(2001).
\bibitem{huang} M.Huang, P.Zhuang, and W.Chao, Phys. Rev. {\bf D 67}, 065015(2003).
\bibitem{fayazbakhsh2} Sh.Fayazbakhsh and N.Sadooghi, Phys. Rev. {\bf D 82}, 045010(2010).
\bibitem{klevansky} S.P.Klevansky, Rev. Mod. Phys. {\bf 64}, 649(1992).
\bibitem{itzykson} C.Itzykson and J.-B Zuber, Quantum Field Theory (McGraw-Hill, New York) (1980).
\bibitem{boomsma} J.K.Boomsma and D.Boer, {\bf D 81}, 074005 (2010).
\end{thebibliography}
\end{document}